\documentclass[pra,twocolumn,floatfix]{revtex4}
\usepackage{graphicx}
\usepackage{amsmath}
\newcommand{\noverk}[2]{\ensuremath{{#1 \choose #2} }}

\begin{document}

\title{Density correlations in ultracold Fermi systems within the
  exact Richardson solution}

\author{Simon Staudenmayer}
\author{Wolfgang Belzig} \email{Wolfgang.Belzig@uni-konstanz.de}
\affiliation{Department of Physics, University of Konstanz, D-78457
  Konstanz, Germany} 

\author{C. Bruder} 
\affiliation{Department of Physics, University of Basel, 
Klingelbergstrasse 82, CH-4056 Basel, Switzerland}

\begin{abstract}
  We discuss the occupation number correlations in an ultracold system
  of interacting fermionic atoms. For a system with a special energy-level
  distribution, viz. two multiply-degenerate levels, explicit
  expressions for the correlation functions are derived in a canonical
  approach using the exact ground state wavefunction of the reduced
  BCS Hamiltonian. We evaluate the correlators numerically for different
  interaction strength and find analytical expressions in
  some limiting cases. Due to the underlying fermionic nature of the
  pairs the occupations are predominantly anti-correlated and their
  statistics is a multinomial distribution.
\end{abstract}



\maketitle

\section{Introduction}
\label{sec:level1}

Ultracold fermionic gases have attracted considerable attention in
theoretical and experimental physics recently. This has been
intensified after the experimental successes in creating Bose-Einstein
Condensates (BECs) in fermionic clouds. An important step was the
development of techniques using magnetically detuned Feshbach
resonances \cite{Observ_Reson_Ferm}, which allow to tune the mutual
interaction strength between the fermions over a wide range.  This
novel opportunity to look at a transition from a weakly attractive
Bardeen-Cooper-Schrieffer (BCS) state to a strongly attractive BEC in
one and the same system makes it interesting from a many-body point of
view (see \cite{grimm2007} for a recent review).  Measurements of the
interaction strength of a fermionic gas near a Feshbach resonance were
made by time-of-flight expansion experiments \cite{Bourdel:03}. The
collective excitations showed a strong dependence on the coupling
strength as was shown experimentally \cite{Bartenstein:04,Kinast:04}.
Other experiments observed condensation \cite{Regal:04,Zwierlein:04}
and the spatial correlations \cite{Greiner:04} of the fermionic atom
pairs in the full crossover regime. Using a spectroscopic technique
the pairing gap was measured directly \cite{denschlag:04}. The
remarkable result was that the gap values were found in good
agreement with a simple BCS expression in the whole crossover
regime.

One way to access the nature of the many-body state is to consider its
statistics.  A number of works proposed to use noise and higher-order
correlations to probe the many-body states of ultracold atoms
\cite{lukin:04,budde:04,meiser:04,lamacraft05a,belzig_schroll_bruder}.
For the BEC-BCS transition the density and spin structure factor was
calculated \cite{buechler:04}.  Schemes to measure the spatial pairing
order interferometrically were proposed based on correlations in
different output channels \cite{Carusotto:04}. To look at pairing
fluctuations of trapped Fermi gases has been proposed in
\cite{viverit:04}.  Experimentally in \cite{grimm:04} the spatial
structure of an atomic cloud was observed directly. This enables to
determine the density fluctuations for example by repeating the
experiment many times or by extracting densities at different
positions in a homogeneous system to obtain the statistics.  The shot
noise of an atomic beam has been experimentally investigated both in
bosonic and fermionic systems
\cite{foelling:05,aspect2005,greiner:05,bloch2006,esteve2006,aspect2007,porto2007}.
Further aspects of full counting statistics in ultracold atomic
systems are discussed in the experimental work in
Ref.~\onlinecite{esslinger2005} and the theoretical papers
\cite{lamacraft05b,meystre,moritz,svistunov,galitski:07}.

Recently, Amico and coworkers have considered the exact solution of
the BCS model in some systems using the algebraic Bethe-Ansatz
\cite{amico:xx}. Explicit expressions for average occupations and the
number correlators have been obtained. Subsequent work has tackled
the problem numerically and found the Bethe-Ansatz solution to be
numerically expensive \cite{faribault:07}. The approach to the
occupation number correlators through the Richardson solution, which
we develop below, can lead to a numerically less expensive method in
some cases. A recent review of the limit of large particle numbers of
the Richardson solution can
be found in \onlinecite{dukelsky:04}.

In a previous publication, we have calculated the full statistics of
particle number fluctuations in ultracold fermionic gases using a
grand-canonical approach \cite{belzig_schroll_bruder}. The idea was to
consider a `bin', i.e., a small subsystem of a homogeneous gas which
contains a macroscopic number of particles, such that the surrounding
atomic gas serves as the particle reservoir.  Fluctuations can in
principle be accessed experimentally by performing a series of
measurements of the number of particles in the subsystem at a fixed
interaction constant, or by considering different bins of the system.
The statistics in the whole BCS-BEC crossover is hence obtained if one
sequentially performs such sets of measurements from small to large
interaction constant.  Due to its effective single-particle form one
can calculate correlation functions using the (grand-canonical)
Bardeen-Cooper-Schrieffer (BCS) ground state solution
\cite{Superconductivity}.  It was found that the BCS-BEC transition
yields a crossover in the statistics of the particle number.
Fluctuations around the average particle number are strongly
suppressed on the BCS side and the statistical distribution is
binomial. On the BEC side, fluctuations are strongly enhanced and are
described by a Poissonian statistics.

Since real ultracold gases consist of a finite number of particles,
the grand-canonical approach may be inappropriate.  In this
article we thus focus on particle-number correlations 
obtained from the exact ground state using the methods developed by
Richardson
\cite{Exact_Eigenstates1,Exact_Eigenstates2,Introduction_Richardson,Numerical_Study,Number_Dependence,Application_Isotopes_Lead}.
The Richardson ground state is an eigenstate of the total particle
number operator $\hat{N}=\sum_i\hat{n}_i$ and thus allows a canonical
treatment of the system. Due to the complexity of the Richardson
solution, we restrict our investigations to a simplified level
distribution consisting of two multiply degenerate energy levels. This
allows us to compare explicitly the thermodynamic limit of the exact
ground state and the approximate BCS solution. Some model-independent
properties of the statistics can be obtained in the limiting cases of
vanishing or very strong interaction.

Although this is a toy model, it is relevant for a number of
experimental situations. First, we note that the large degree of
control possible in ultra-cold atomic systems, e.g. by using optical
lattices or atomic chips, will make it possible to produce few-level
systems, which can be loaded with a fixed number of fermions in a
controlled way. Experiments on particle number correlations in such
systems will be described bny the theory developed below.
Our results apply equally well to level configurations, in which only
two groups of levels are relevant. The level spacing within each group
has to be smaller than the interaction constant; the transition
from weak to strong coupling appears for interactions of the order of
the energy spacing between the groups. We would also like to mention,
that the results obtained below for the strongly interacting limit are
valid for any level configuration, in which the maximal level spacing
is smaller than the interaction constant. Hence, we believe that
systems corresponding to the model we study can be experimentally
produced, or, at least, some predictions can be tested in the limiting
case of a strong interaction in an arbitrary level configuration.

\section{Properties of the exact solution}

\subsection{General case}

We start by recalling the basic properties of the Richardson solution
to the reduced Hamiltonian
\cite{Spectros_Ultra_Grain,Introduction_Richardson}. The Hamiltonian
in second quantization and momentum space has the form
\begin{equation}
  H=\sum_f 2\epsilon_f
  \hat{n}_f-g\sum_{ff^{\prime}}\hat{b}_{f}^{\dagger} 
  \hat{b}_{f^{\prime}}^{\phantom{\dagger}}~,
  \label{reduced_hamilt} 
\end{equation}
where $\hat{b}_f=\hat{c}_{f\downarrow}\hat{c}_{f\uparrow}$ are 
hard-core bosonic annihilation operators. The reduced Hamiltonian
captures only the scattering fermions which occur in time-reversed
states. It is therefore possible to express the particle number
operator totally in terms of $\hat{b}$-operators:
\begin{equation}
  \hat{n}_f=\frac{\hat{c}_{f\uparrow}^{\dagger} 
    \hat{c}_{f\uparrow}^{\phantom{\dagger}}
    +\hat{c}_{f\downarrow}^{\dagger}
    \hat{c}_{f\downarrow}^{\phantom{\dagger}}}{2}
  =\hat{b}_f^{\dagger}\hat{b}_f^{\phantom{\dagger}}~.
\end{equation}
This is true only in the subspace of fully paired states, to which we
will restrict ourselves here and in the following.
The $J$-th (with $J=1$:~ground state, $J=2$:~first excited state,
etc.)  $N$-particle eigenstate of (\ref{reduced_hamilt}) has the form
\begin{equation}
  |\Psi_N^{(J)}\rangle=\sum_{f_1\ldots f_N}\varphi^{(J)}(f_1,\ldots, f_N)
\prod_{\nu=1}^Nb_{f_\nu}^{\dagger}|0\rangle~.
\end{equation}
Because of $(b_{f}^{\dagger})^2=0$ only those terms contribute to the
sum for which all of the indices $f_1,\ldots, f_N$ are distinct.  The
coefficient is given by
\begin{equation}
  \varphi^{(J)}(f_1,\ldots, f_N)=
  C\sum_P\prod_{\nu=1}^N\frac{1}
  {2\epsilon_{f_\nu}^{\phantom{J}}-E_{P(\nu)}^{(J)}}\label{varphi}~,
\end{equation}
where $\sum_P$ denotes the sum over all permutations $P(i)$.  The
normalization constant $C$ can be determined applying a standard
determinant method \cite{Exact_Eigenstates2}.  The quasi-energies
$E_\nu^{(J)}$ in Eq.~(\ref{varphi}) are the solutions of the set of
coupled root equations
\begin{equation}
  1-\sum_f^U\frac{g}{2\epsilon_f-E_\nu}+\sum_{\nu^\prime\neq \nu}^{N}
  \frac{2g}{E_{\nu^\prime}-E_\nu}=0,\quad \nu=1\ldots
  N~.\label{quasi_energies} 
\end{equation}
In general, the $E_\nu^{(J)}$ are complex quantities, however they
always appear in complex-conjugate pairs.  The corresponding energy
eigenvalue is given by
\begin{equation}
  \varepsilon_N^{(J)}=\sum_{\nu=1}^{N}E_\nu^{(J)}~,\label{total_energy}
\end{equation}
and is thus real, as required.

\subsection{The two-level model}

We will now consider the special configuration involving $N$ particles
in two multiply-degenerate energy levels $\epsilon_0$ and $\epsilon_1$
with the degeneracies $\Omega_0$ and $\Omega_1$, respectively .  In
the following the subscripts $0$ and $1$ will refer to one of the
lower levels and one of the upper levels, respectively.  The
coefficients $\varphi$ from Eq.~(\ref{varphi}) can be expressed in
terms of a single variable $\nu$ 
that indicates the amount of particles in the upper level $\epsilon_1$
\cite{Exact_Eigenstates1}:
$\varphi^{(J)}(f_1,\ldots, f_N)\to\varphi^{(J)}(\nu)$.
This leads to a simplification in finding the Richardson solution.
Introducing the abbreviations
\begin{eqnarray}
  \omega_{\nu}&=&2N\epsilon_0+2\nu (\epsilon_1-\epsilon_0)-
g\nu(\Omega_1-\nu+1)\nonumber\\
  &&- g(N-\nu)(\Omega_0-N+\nu+1)~,\\
  A_{\nu}&=&g(N-\nu)(\Omega_1-\nu)~,\\
  B_{\nu}&=&g\nu(\Omega_0-N+\nu)~,
\end{eqnarray}
the coefficients are determined by the continued-fraction formula
\begin{equation}
  \varphi^{(J)}(\nu)=\cfrac{B_{\nu}\varphi^{(J)}(\nu-1)}{\omega_{\nu}-
    \varepsilon_N^{(J)}-\cfrac{A_{\nu}B_{\nu+1}}{\omega_{\nu+1}-
      \varepsilon_N^{(J)}-\ldots{{}
        -\cfrac{A_{N-1}B_{N}}{\omega_{N}-\varepsilon_N^{(J)}}}}}\,.
  \label{final_deteq_for_phi_in_2level_model}
\end{equation}
$\varphi^{(J)}(0)$ has to be extracted from the normalization condition
\begin{equation}
  \sum_{\nu=0}^{N}\noverk{\Omega_0}{N-\nu}\noverk{\Omega_1}{\nu}
  \big[{\varphi^{(J)}}(\nu)\big]^2=1\;.
\label{norm_cond}
\end{equation}
One can find an expression for the total energy (\ref{total_energy}) appearing in
Eq.~(\ref{final_deteq_for_phi_in_2level_model}) directly from the root
equation
\begin{eqnarray}
  \omega_{0}-\varepsilon_N=-\cfrac{A_{0}B_{1}}{\omega_{1}-
    \varepsilon_N-\cfrac{A_{1}B_{2}}{\omega_{2}-\varepsilon_N-\ldots{{}
        -\cfrac{A_{N-1}B_{N}}{\omega_{N}-\varepsilon_N}}}}\,,
\label{root_eq_2levelmodel}
\end{eqnarray}
without having to resort to the quasi-energies from
Eqs.~(\ref{quasi_energies}). 

In the following, we will use the level spacing $\epsilon_1-\epsilon_0=1$
as our energy scale. The only (dimensionless) parameter left to
characterize the system is the ratio between the interaction constant
$G$ and the level spacing.

The model of two highly-degenerate levels is clearly an artificial
model, which cannot be mapped to generic many-particle
systems. However, we believe the model is nonetheless relevant also
for experimental realizations due to two reasons. First, we show below
that deviations of the occupation number correlators from a simple BCS
mean-field treatment are relevant and can be detected in not too large
systems. In experimental realizations of interacting Fermi systems in
ultracold atomic gases an unprecedented variability of system
parameters has been experimentally demonstrated \cite{grimm2007}.  We
therefore hope that our investigation will stimulate experimental
efforts to create few-particle strongly-interacting systems and study
the effects we predict below. Second, within the same reasoning we
believe that the large variability of tailoring atom systems in
(magneto-)optical traps or via atomic chips will make it possible to
create an artificial highly-degenerate two-level system and to use it
to study in a controlled manner the transition to the thermodynamic
limit in a particularly simple system, as we predict here.

\subsection{Example}
\label{sec:example}

\begin{figure}
  \includegraphics[width=\columnwidth,clip=true]{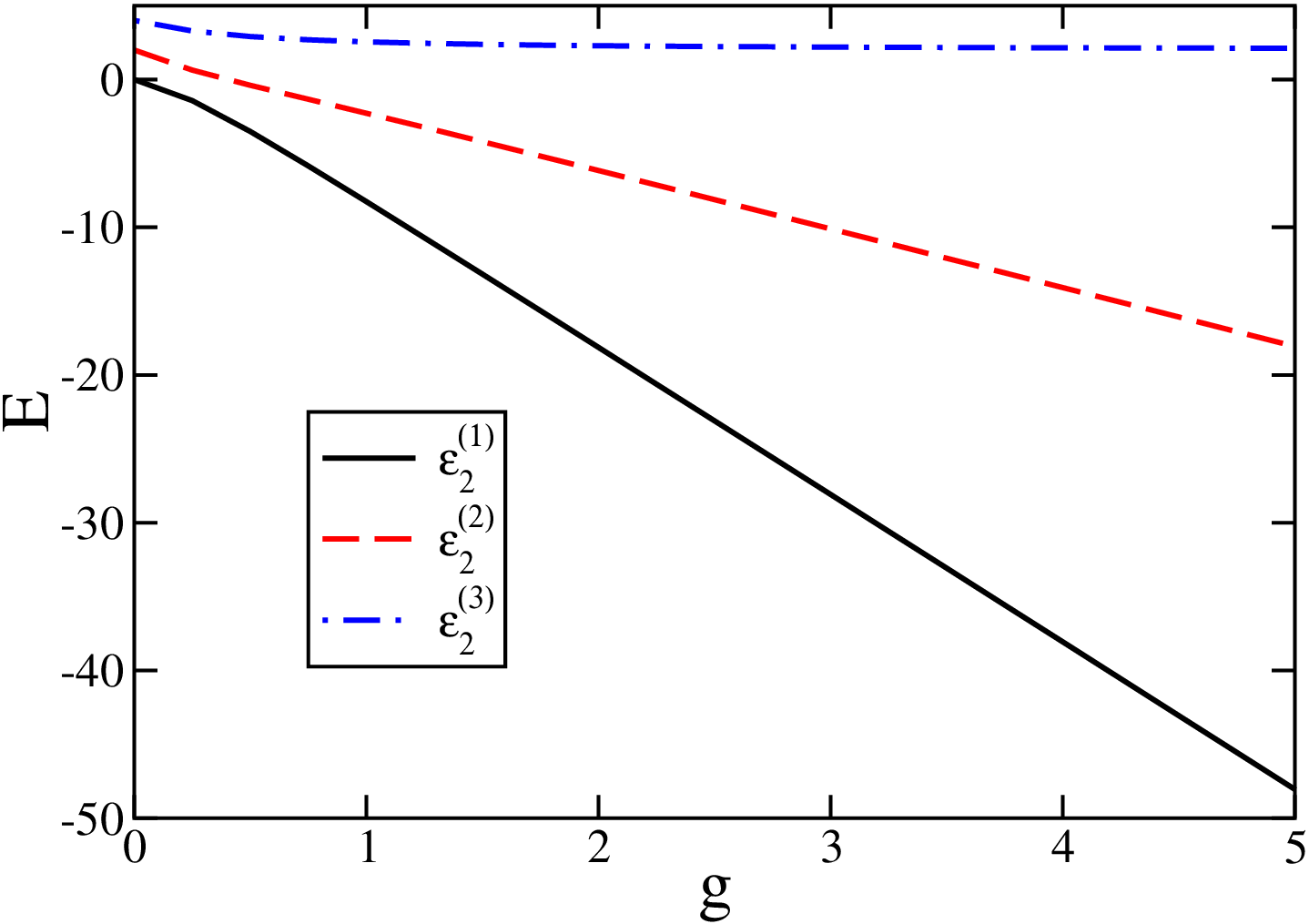}
\caption{\label{fig:energ} (Color online) Development of the ground
  state and first and second excited state energy vs. interaction
  constant $g$ for a simple two-boson system. The two levels at
  $\epsilon_0=0$ and $\epsilon_1=1$ are both threefold degenerate.}
\end{figure}

We want to illustrate some characteristics of the Richardson solution
by means of a simple setup within the two-level model. We consider,
therefore, a system of two hard-core bosons in two threefold degenerate
levels ($\Omega_0=\Omega_1=3$) of energy $\epsilon_0=0$ and
$\epsilon_1=1$. Figure~\ref{fig:energ} shows the three root solutions
of Eq.~(\ref{root_eq_2levelmodel}) as a function of the interaction
$g$.  Obviously, in the non-interacting limit at $g\ll 1$, the
energies reduce to the bare pair energies 0, 2 and 4,
corresponding to the case that $\nu=0,~\nu=1$ and $\nu=2$ particles
respectively are in the upper level. With increasing interaction
$g\approx 1$, the ground state and first excited state energies are
lowered continuously, whereas the second excitation energy
approaches $\varepsilon_2^{(3)}=2$ and is then independent of the
interaction constant $g$.

\begin{figure}
  \includegraphics[width=\columnwidth,clip=true]{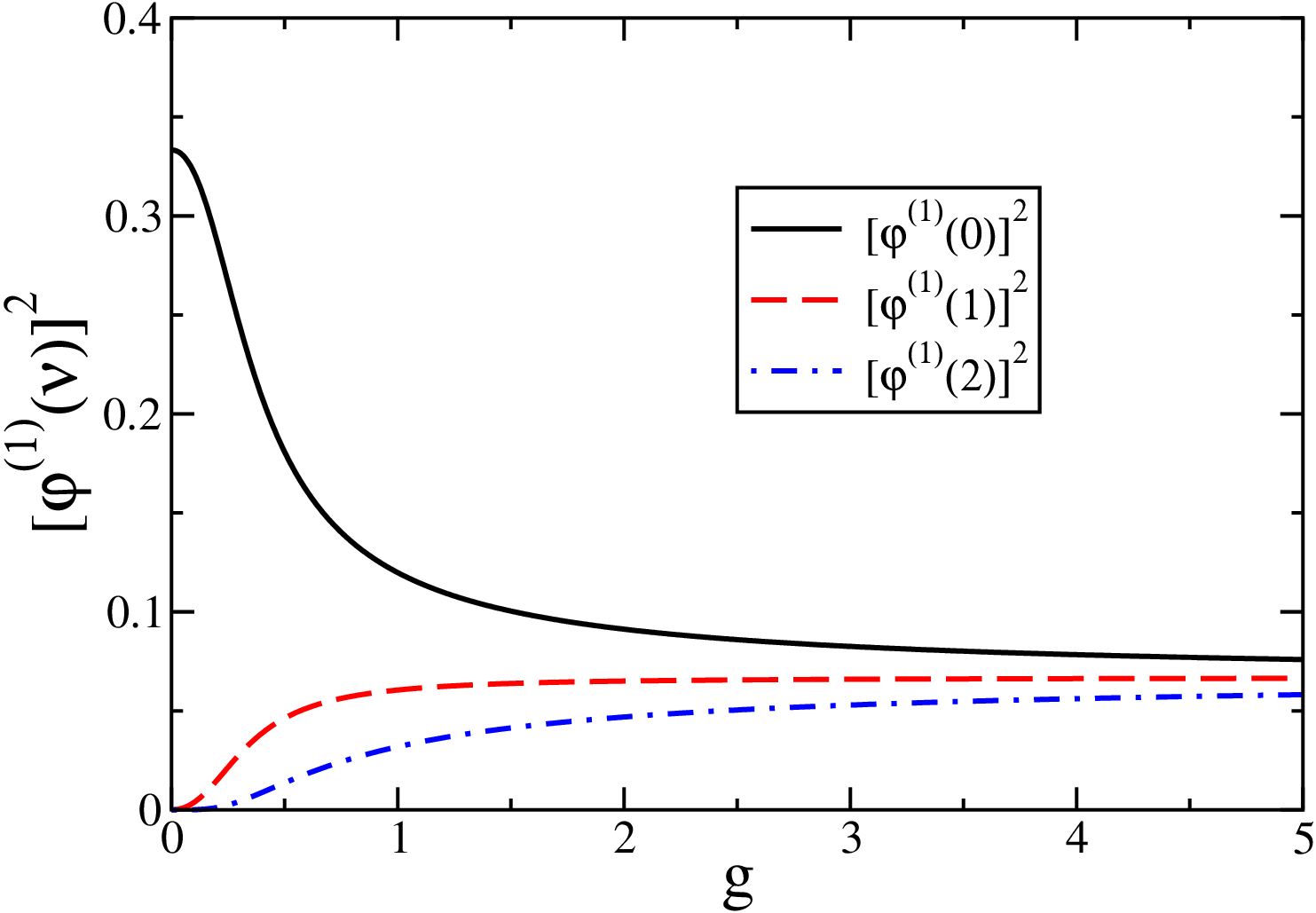}
  \caption{\label{fig:Coeff_2_3E1} Many-body coefficient
    $[\varphi^{(1)}(\nu)]^2$ as a function of $g$ for the
    ground state energy $\varepsilon_2^{(1)}$.} 
\end{figure}

In the following, we concentrate our investigations on the behavior of
the ground state.  Figure~\ref{fig:Coeff_2_3E1} shows the behavior of
the many-body occupation $\big[\varphi^{(1)}(\nu)\big]^2$ as a
function of $g$.  At vanishing interaction only the lower three energy
levels at $\epsilon_0$ are occupied. From the normalization condition
(\ref{norm_cond}) it thus follows that $[\varphi^{(1)}(0)]^2 = 1/3$,
since there are $\noverk{3}{2}=3$ distinct possibilities to distribute
two particles among three levels. The average occupation number of a
lower level is hence given by $\langle\hat{n}_0\rangle = \frac{2}{3}$.
At strong interactions $g\gg 1$, all levels tend to become equally
occupied. In this limit, we can therefore neglect the level spacing
and consider simply a single energy level with a total a single total
degeneracy $\Omega=\Omega_0+\Omega_1=6$.  Equation~(\ref{norm_cond})
simplifies to $[\varphi^{(1)}(\nu)]^{-2} = \noverk{\Omega}{N} =
\noverk{6}{2} = 15$, which is independent of $\nu$, and the average
particle number of a level is given by $\langle\hat{n}_0\rangle =
\langle\hat{n}_1\rangle = N/\Omega = 1/3$. We want to point out here
that the equal occupation of all levels in the strong
interacting limit is not restricted to this specific level model
but rather a general feature of the Richardson ground state.

For $g\gg 1$ it follows from
Eq.~(\ref{final_deteq_for_phi_in_2level_model}) and
Eq.~(\ref{root_eq_2levelmodel}) that the ground state energy
approaches
\begin{eqnarray}
  \varepsilon_N^{(1)}=\omega_0-A_0\frac{\varphi^{(1)}(1)}{\varphi^{(1)}(0)} 
  \approx 2N\epsilon_0-gN(\Omega-N+1)\,,
\end{eqnarray}
since $\varphi^{(1)}(1)\approx \varphi^{(1)}(0)$.  This useful relation can e.~g.
be taken as an initial energy guess in the whole interaction regime,
when it comes to finding the roots of
Eq.~(\ref{root_eq_2levelmodel}).

\section{particle-number correlations: grand-canonical vs. canonical}

We address now the correlations following from the exact ground state.
We particularly focus on the differences that occur in a canonical
treatment compared to applying the grand-canonical BCS solution.  At
first, we define the particle number cross-correlator between the
occupations of levels $f\neq f^\prime$
\begin{eqnarray}
  g(f,f^\prime) &:=& \langle(\hat{n}_f-\langle\hat{n}_f
  \rangle)(\hat{n}_{f^\prime}-\langle\hat{n}_{f^\prime}\rangle)\rangle
\nonumber\\
&=&\langle\hat{n}_f\hat{n}_{f^\prime}\rangle-\langle\hat{n}_f
  \rangle\langle\hat{n}_{f^\prime}\rangle\,,
  \label{PNFCC}
\end{eqnarray}
which represents a direct measure of how much the particle number of
level $f$ fluctuates around its mean value in the presence of a
fluctuation around the mean value of the particle number of level
$f^\prime$.  

The grand-canonical BCS wavefunction \cite{Superconductivity} is given
by
\begin{equation}
  |\textnormal{BCS}\rangle=\prod_f\big(u_f+v_fb_f^{\dagger}\big)|0\rangle~,
\label{BCS_groundstate}
\end{equation}
with
$v_f^2=1-u_f^2=(1-(\epsilon_f-\mu)/\sqrt{(\epsilon_f-\mu)^2+\Delta^2})/2$,
where $\mu$ is the chemical potential. The mean field $\Delta$ and
$\mu$ are fixed by the self-consistency equations
\begin{equation}
  \Delta=-g\sum_fu_fv_f~,~\bar{N}=2\sum_fv_f^2~.\label{self_cons_eqns}
\end{equation}
The simplification by the mean field Ansatz, that is the reduction of
the many-body interaction to an effective one-body interaction, has a
direct consequence on cross-correlations: Since 
(\ref{BCS_groundstate}) is a product state the different level
occupations are uncorrelated and, hence, $g(f,f^\prime)=0$.  As
we will see in the following sections, the correlations will be
non-zero if the many-body interaction is taken into account beyond the
mean-field approach.

Due to the operator identity $\hat{n}_f=\hat{n}_f^l$ for $l=1,2,...$
in the subspace of paired particles, the auto-correlation function of a level,
Eq.~(\ref{PNFCC}) with $f=f^\prime$, is totally determined by its average
particle number and thus does not contain any additional information.
In the following, we will therefore concentrate on the investigation
of exact average particle numbers and exact particle number
cross-correlators in the form of Eq.~(\ref{PNFCC}).

\subsection{Exact correlators in the two-level model}

We now determine the explicit form of the particle number
cross-correlator (\ref{PNFCC}) in the two-level model. We only have to
consider three different kinds of correlators, since all degenerate
levels are equivalent. If two levels of the same energy are distinct,
we indicate this by priming one of the indices labeling the energy of
the level. The three different cases take the form 
\begin{eqnarray}
  g(0,0^{\prime}) & = & \langle\hat{n}_0\hat{n}_{0^{\prime}}\rangle-
  \langle\hat{n}_0\rangle^2\label{corr_ll}\\
  g(0,1) & = & \langle\hat{n}_0\hat{n}_{1}\rangle-
  \langle\hat{n}_0\rangle\langle\hat{n}_1\rangle \label{corr_lu}\\
  g(1,1^{\prime}) & = & \langle\hat{n}_1\hat{n}_{1^{\prime}}\rangle-
  \langle\hat{n}_1\rangle^2\label{corr_uu}
\end{eqnarray}
with (assuming that $N\leq\Omega_0,~\Omega_1$) 
\begin{eqnarray}
  \langle\hat{n}_0\hat{n}_{0^{\prime}}\rangle & = & \sum_{\nu=0}^{N-2}
  \noverk{\Omega_0-2}{N-2-\nu}\noverk{\Omega_1}{\nu}
  \left[\varphi^{(1)}(\nu)\right]^2\,,\label{corr_ll_two_level_model}
  \\\nonumber
  \langle\hat{n}_0\hat{n}_1\rangle & = & \sum_{\nu=1}^{N-1} 
  \noverk{\Omega_0-1}{N-1-\nu}\noverk{\Omega_1-1}{\nu-1} 
  \left[\varphi^{(1)}(\nu)\right]^2\,,
  \\\nonumber
  \langle\hat{n}_1\hat{n}_{1^{\prime}}\rangle & = & \sum_{\nu=2}^{N} 
  \noverk{\Omega_0}{N-\nu}\noverk{\Omega_1-2}{\nu-2} 
  \left[\varphi^{(1)}(\nu)\right]^2
\end{eqnarray}

and
\begin{eqnarray}
  \langle\hat{n}_0\rangle & = & \sum_{\nu=1}^{N}
  \noverk{\Omega_0-1}{N-1-\nu}\noverk{\Omega_1}{\nu}
  \left[\varphi^{(1)}(\nu)\right]^2\,,\label{averagePN_l_two_level_model}
  \\
  \langle\hat{n}_1\rangle & = & \sum_{\nu=0}^{N-1} 
  \noverk{\Omega_0}{N-\nu} \noverk{\Omega_1-1}{\nu-1}
  \left[\varphi^{(1)}(\nu)\right]^2\,.\label{averagePN_u_two_level_model}
\end{eqnarray}

If we allow particle numbers exceeding one or both degeneracies, the
boundaries in the sums appearing in Eqs.~(\ref{norm_cond}),
(\ref{corr_ll_two_level_model}) and
(\ref{averagePN_l_two_level_model}) have to be adjusted accordingly.

\subsection{Relation to counting statistics}

In the two-level model, we are also able to specify the full
statistics of the occupation numbers.  This quantity can in principle
be obtained by measuring repeatedly the occupation numbers and finding
the probability $P(n_f,n_{f^\prime})$ that two levels $f$ and
$f^\prime$ have occupations $n_f$ and $n_{f^\prime}$. This full
statistics can be equivalently expressed through the cumulant
generating function $S_{f{f^\prime}}(\chi_f,\chi_{f^\prime})=\ln
\sum_{n_f,n_{f^\prime}} \exp(i\chi_fn_f+i\chi_{f^\prime} n_{f^\prime})
P_{f{f^\prime}}(n_f,n_{f^\prime})$ \cite{belzig_schroll_bruder}.  The
cumulant generating function for hard-core bosons in a fully paired
state is given as a function of two counting fields $\chi_f$ and
$\chi_{f^\prime}$:
\begin{eqnarray}
  e^{S_{f{f^\prime}}(\chi_f,\chi_{f^\prime})}&=&
  \langle e^{i(\chi_f\hat{n}_f+\chi_{f^\prime}\hat{n}_{f^\prime})}\rangle
  \label{CGF}
  \\ 
  & = & 1+\langle\hat{n}_f\rangle\left(e^{i\chi_f}-1\right)
  \nonumber\\ & & +
  \langle\hat{n}_{f^\prime}\rangle\left(e^{i\chi_{f^\prime}}-1\right)
  \nonumber\\\nonumber
  &&+\langle\hat{n}_f\hat{n}_{f^\prime}\rangle 
  \left(e^{i\chi_f}-1\right)\left(e^{i\chi_{f^\prime}}-1\right)\,.
\end{eqnarray}
Consequently, the only correlator which needs to be known to fully
determine the CGF is the one in the last line of Eq.~(\ref{CGF}), for
which we are able to give explicit expressions here, due to the
simplicity of the model.

In the case of non-interacting particles, e.g. hard-core bosons in the BCS
mean-field treatment, Eq.~(\ref{CGF}) factorizes according to
\begin{eqnarray}
  e^{S_{ff^\prime}(\chi_f,\chi_{f^\prime})} & = & 
  e^{S_f(\chi_f)+S_{f^\prime}(\chi_{f^\prime})} \\\nonumber 
  & = & \left[1+\langle\hat{n}_f\rangle\left(e^{i\chi_f}-1\right)\right]
  \\\nonumber
  & & \times
  \left[1+\langle\hat{n}_{f^\prime}\rangle
  \left(e^{i\chi_{f^\prime}}-1\right)\right]\,.
\end{eqnarray}
This is the CGF of uncorrelated particle numbers. Comparing these general
results for the counting statistics with the correlators discussed in
the previous subsection we observe that in this special case the
counting statistics contains not more information than the correlators
alone. Or, in other words, if the correlators Eq.~(\ref{corr_ll})-(\ref{corr_uu}) are
known, one can use Eq.~(\ref{CGF}) to calculate the full counting
statistics.

\subsection{Asymptotic behavior}

Before we discuss the general results for an arbitrary interaction
constant, we obtain analytical expressions for the correlators in the
limiting cases of weak and strong interactions. This is possible since
the coefficients (\ref{varphi}) can be directly determined from the
normalization condition (\ref{norm_cond}) without having to solve the
root equation, Eq.~(\ref{root_eq_2levelmodel}).  In the following, we will
assume that $N\leq\Omega_0,~\Omega_1$.  For $g\ll 1$,
Eq.~(\ref{norm_cond}) reduces to
$\left[\varphi^{(1)}(0)\right]^{-2}=\noverk{\Omega_0}{N}$. From
Eq.~(\ref{corr_ll_two_level_model}) and
Eq.~(\ref{averagePN_l_two_level_model}) thus follows
\begin{equation}
  \label{eq:average_gll_asymptotic}
  g(0,0^{\prime})=
  -\langle\hat{n}_0\rangle^2\frac{1-\langle\hat{n}_0\rangle}
  {N-\langle\hat{n}_0\rangle}\,,
\end{equation}
with the system-size-independent average particle number
$\langle\hat{n}_0\rangle=N/\Omega_0$. The remaining
correlators are zero. 

For $g\gg 1$ we have correspondingly
$\left[\varphi^{(1)}(\nu)\right]^{-2}=\noverk{\Omega}{N}$, where
$\Omega=\Omega_0+\Omega_1$. Hence we get
\begin{equation}
  g(f,{f^\prime})=
  -\langle\hat{n}\rangle^2\frac{1-\langle\hat{n}\rangle
  }{N-\langle\hat{n}\rangle}\,.
  \label{PNFCC_strong_G}
\end{equation}
Again $\langle\hat{n}\rangle=N/\Omega$ is the system-size and
interaction-constant independent
average occupation number of every level.  Since it is a feature of the
Richardson solution, that all coefficients $\big[\varphi^{(1)}(f_1\ldots
f_N)\big]^2$ become equal in the strongly-interacting limit,
(\ref{PNFCC_strong_G}) is a universal property of particle-number
correlators, which is valid also for arbitrary level configurations
and not only restricted to this simple model.

\subsection{The two-level model at half filling}

We now discuss the numerical results of the average particle numbers
and correlators above as we approach the thermodynamic limit starting
from finite system sizes. In the evaluation of e.~g.
Eq.~(\ref{corr_ll_two_level_model}) and
Eq.~(\ref{averagePN_l_two_level_model}), we hence have to assure that
the involved quantities scale in the correct manner. 
The continuum
limit is obtained by taking $\Omega\to\infty$, while leaving
$\frac{N}{\Omega}$ and $G=g \Omega$ constant
\cite{Etats_Propres,Pairing_Limit,Large_N_Limit}. We call $G$ the
`system-size-independent coupling constant'. Under these assumptions,
increasing the particle number will lead to the BCS results in the
thermodynamic limit. 

\begin{figure}
  \includegraphics[width=\columnwidth,clip=true]{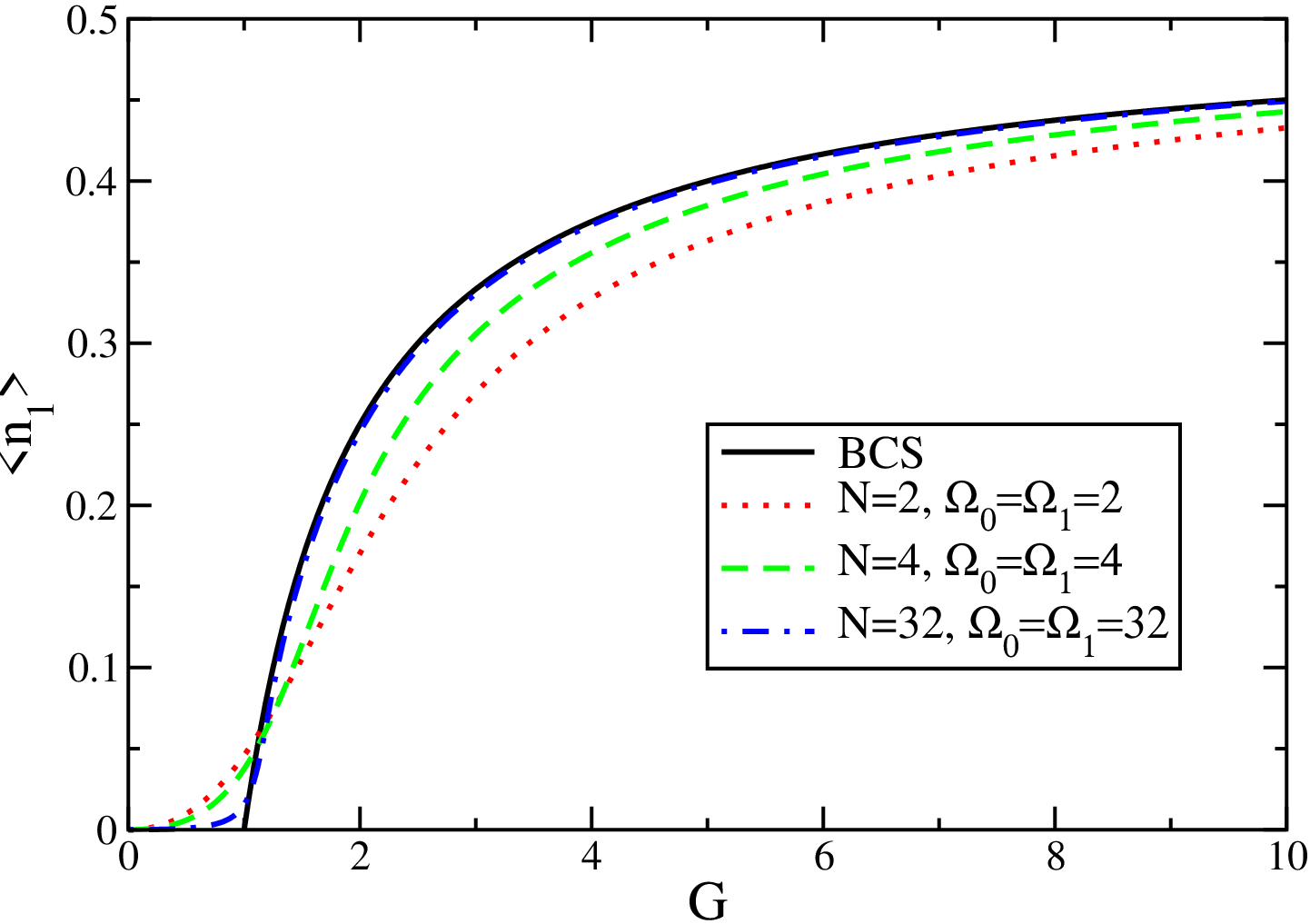}\\
  \caption{\label{fig:occup} (Color online) Average particle number of
    one of the upper levels as
    a function of $G$ for various system sizes. The black solid lines
    correspond to the solutions obtained from BCS theory. Note that
    the self-consistency Eqs.~(\ref{self_cons_eqns}) in this case only
    have real solutions for $G>1$. The average particle number of one
    of the lower levels follows from 
$\langle\hat{n}_0\rangle=1-\langle\hat{n}_1\rangle$.}
\end{figure}

At first, we investigate the two-level model at half filling with
equal degeneracies of both energy levels, viz.~$\Omega_0=\Omega_1=N$.
Figure~\ref{fig:occup} shows the average particle number in one of the
upper levels as a function of $G$ for various system sizes.  We can
see that there is already a fairly good agreement to the BCS results
in the case of only 32 particles.
Due to the particle-hole symmetry of the system,
the connection between the average particle number of a lower
level and an upper level is given by 
$\langle\hat{n}_0\rangle=1-\langle\hat{n}_1\rangle$. 
The average particle numbers
in the limits of weak and very strong interactions are system-size
independent: for $G\ll 1$, only the lower energy band is occupied. For
$G\gg 1$ as mentioned in Sec.~\ref{sec:example}, we obtain an equal
occupation of all levels. 

\begin{figure}
  \includegraphics[width=\columnwidth,clip=true]{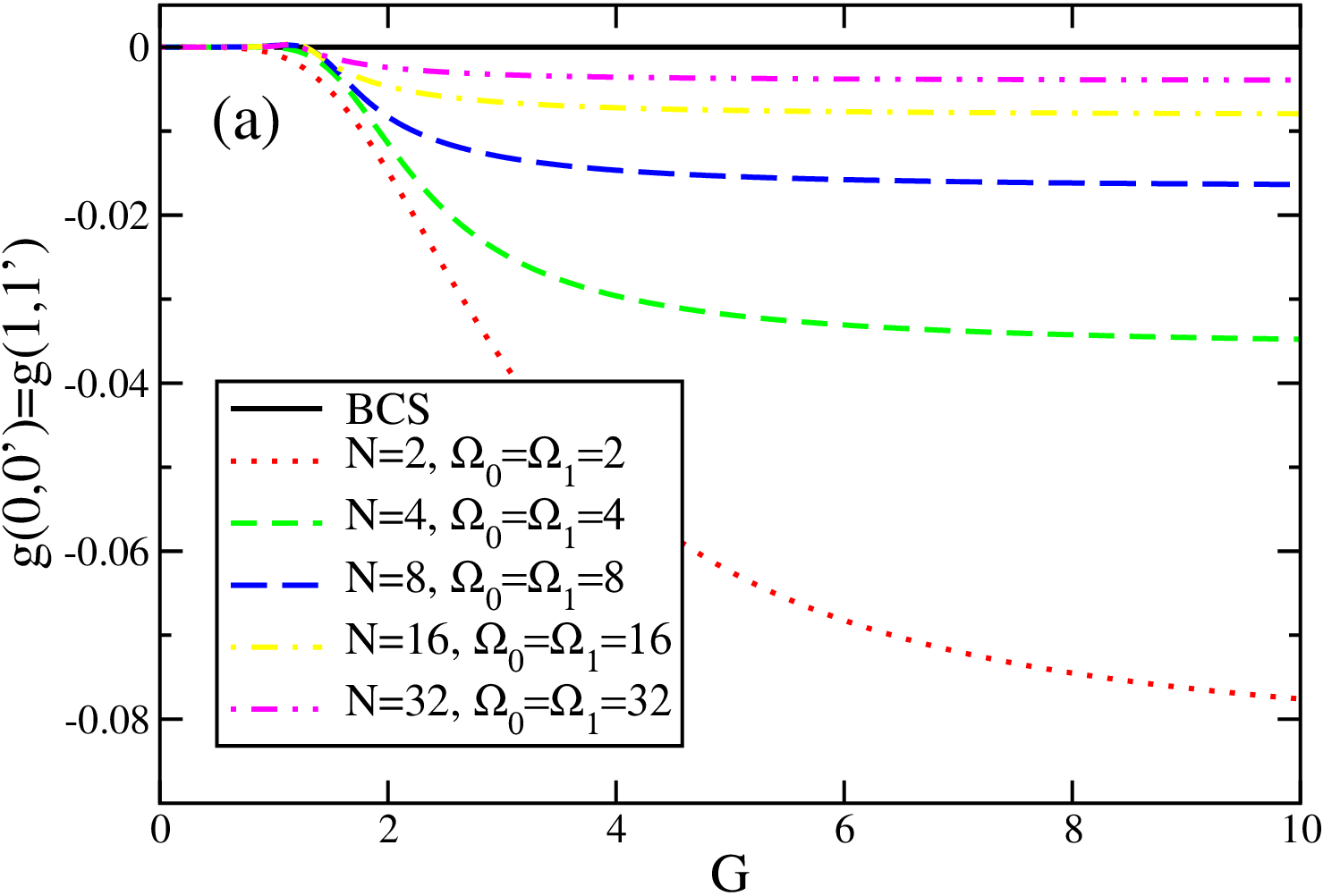}\\
  \includegraphics[width=\columnwidth,clip=true]{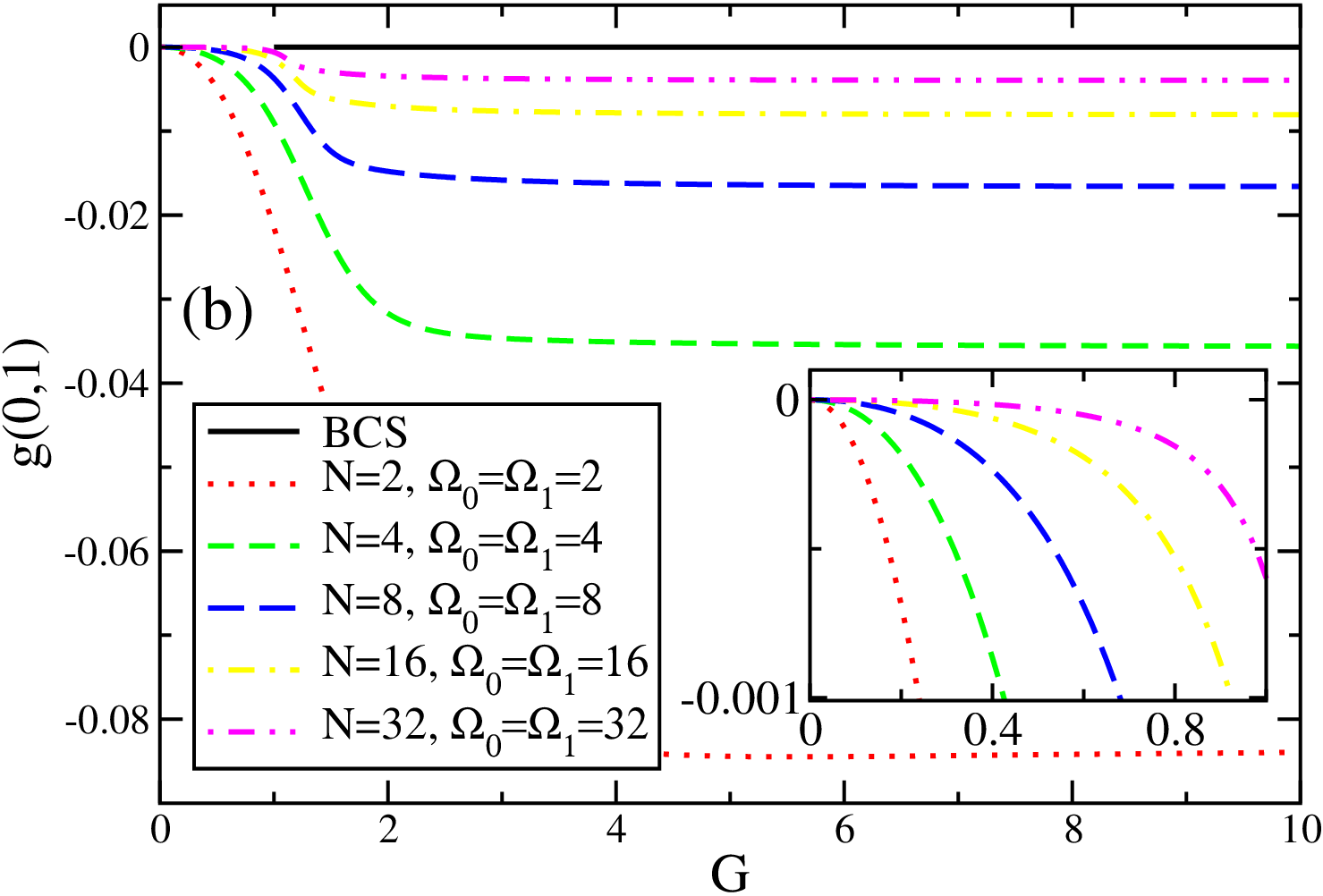}
  \caption{(Color online)
  $g(0,0^{\prime})=g(1,1^{\prime})$ (upper plot) and $g(0,1)$ (lower plot) as a
  function of $G$ for various system sizes. The inset shows that, in
  contrast to $g(0,0^{\prime})$, $g(0,1)$ is always negative in the 
low-interaction regime.} 
\label{fig:corr_HF} 
\end{figure}

In Fig.~\ref{fig:corr_HF} the corresponding correlations are given as
a function of $G$. Note that, due to particle-hole symmetry in the
half-filled case $g(0,0^{\prime}) = g(1,1^{\prime})$.  The behavior
of the average particle numbers in the strongly-interacting case has a
direct influence on the correlations causing $g(0,0^{\prime})$,
$g(0,1)$ and $g(1,1^{\prime})$ to become equal in magnitude for a
fixed system size. At vanishing interaction, the lacking possibility
of reshuffling particles in a fully occupied band leads to
zero-correlation. A comparison of the plots in Fig.~\ref{fig:corr_HF}
shows that for a given $N$, the crossover happens over a smaller range
of $G$ in the case of $g(0,1)$.  Obviously, the fact that all
occupations $\nu$ contribute to this correlator - contrary to
Eq.~(\ref{corr_ll}), where $\nu=N$ only enters through the
normalization (\ref{norm_cond}) - leads to a faster saturation with
increasing coupling constant.

\begin{figure}
  \includegraphics[width=\columnwidth,clip=true]{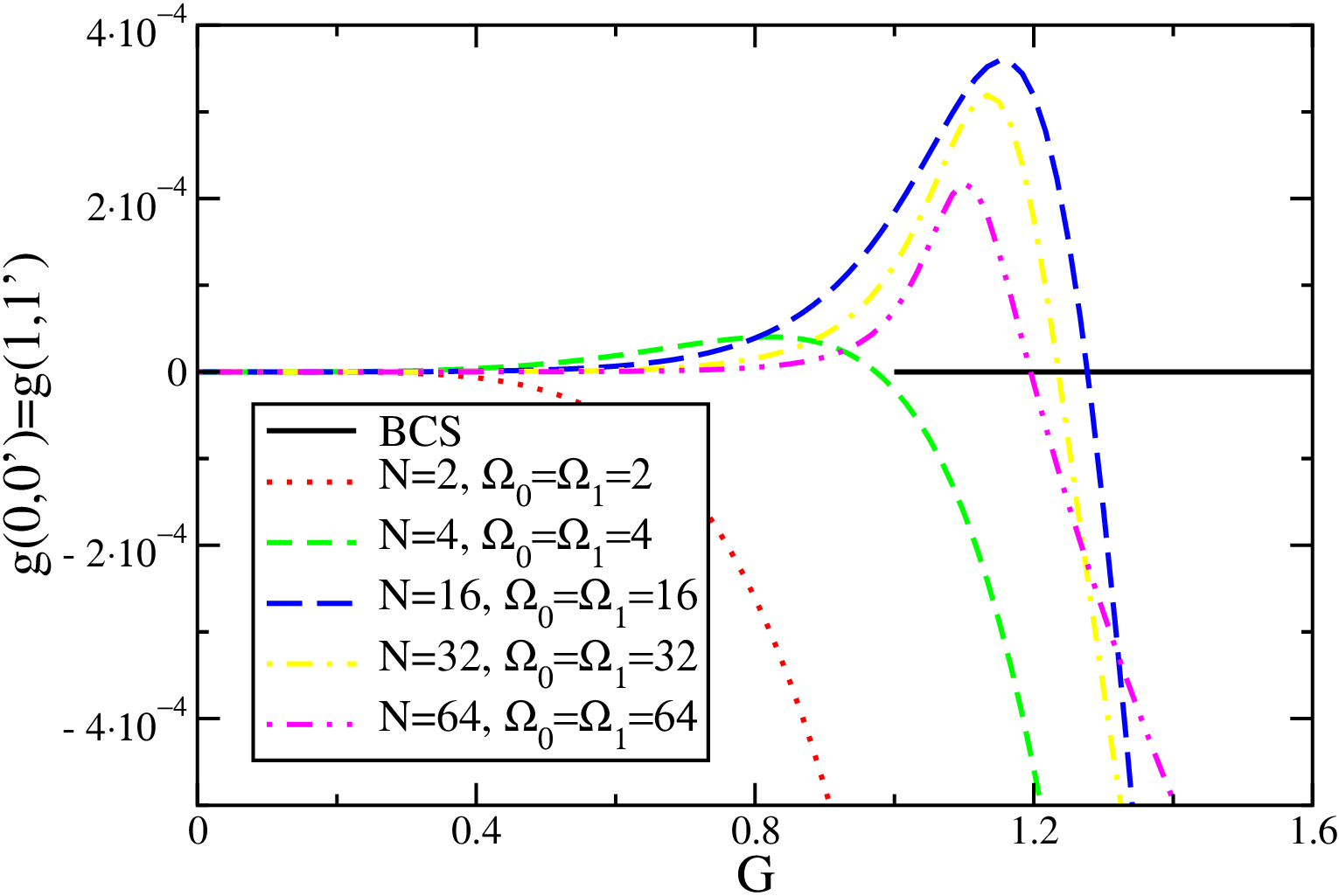}
  \caption{\label{fig:corr_HF_higher_res} (Color online)
    Zoom into the region of positive correlations of
    $g(0,0^{\prime})=g(1,1^{\prime})$ in the weak-coupling range for
    different system sizes. A positive peak develops around $G=1$ for
    small systems and becomes maximal for $N\approx 16$. For large
    system sizes the absolute values become smaller again, but the
    overall feature sharpens.}
\end{figure}

As a general feature, one finds that the particle number correlators
of distinct levels tend to converge to the zero-correlation line of
the mean-field approach in the whole interaction regime as one
increases the number of particles. A direct indication of the
fermionic origin of the hard-core bosons is that, at first sight, in
the non-limiting cases $N\neq\infty$ and $G\neq 0$ all correlators are
negative, corresponding to anti-correlated particle numbers: Due to
the presence of a particle in level $f$, it is less probable to find
another particle at the same time in level ${f^\prime}$ than in the
uncorrelated case. However for $g(0,0^{\prime})$, we observe a range
at intermediate interactions, where particles of a certain energy
promote other particles to occupy the same level.  It also leads to
another point of vanishing correlation for $G\neq 0$ and fixed $N$;
see Fig.~\ref{fig:corr_HF_higher_res}. Evaluating $g(0,0^{\prime})$ in
second-order perturbation theory shows that this effect starts to
occur for $N>2$. There is a resonance effect with a maximum peak value
in the positive correlation between 16 and 32 particles. We do not
find a positive range for $g(0,1)$. This surprising finding is
confirmed analytically by a perturbative calculation in the appendix.

\subsection{The two-level model away from half filling}

\begin{figure}
  \includegraphics[width=\columnwidth,clip=true]{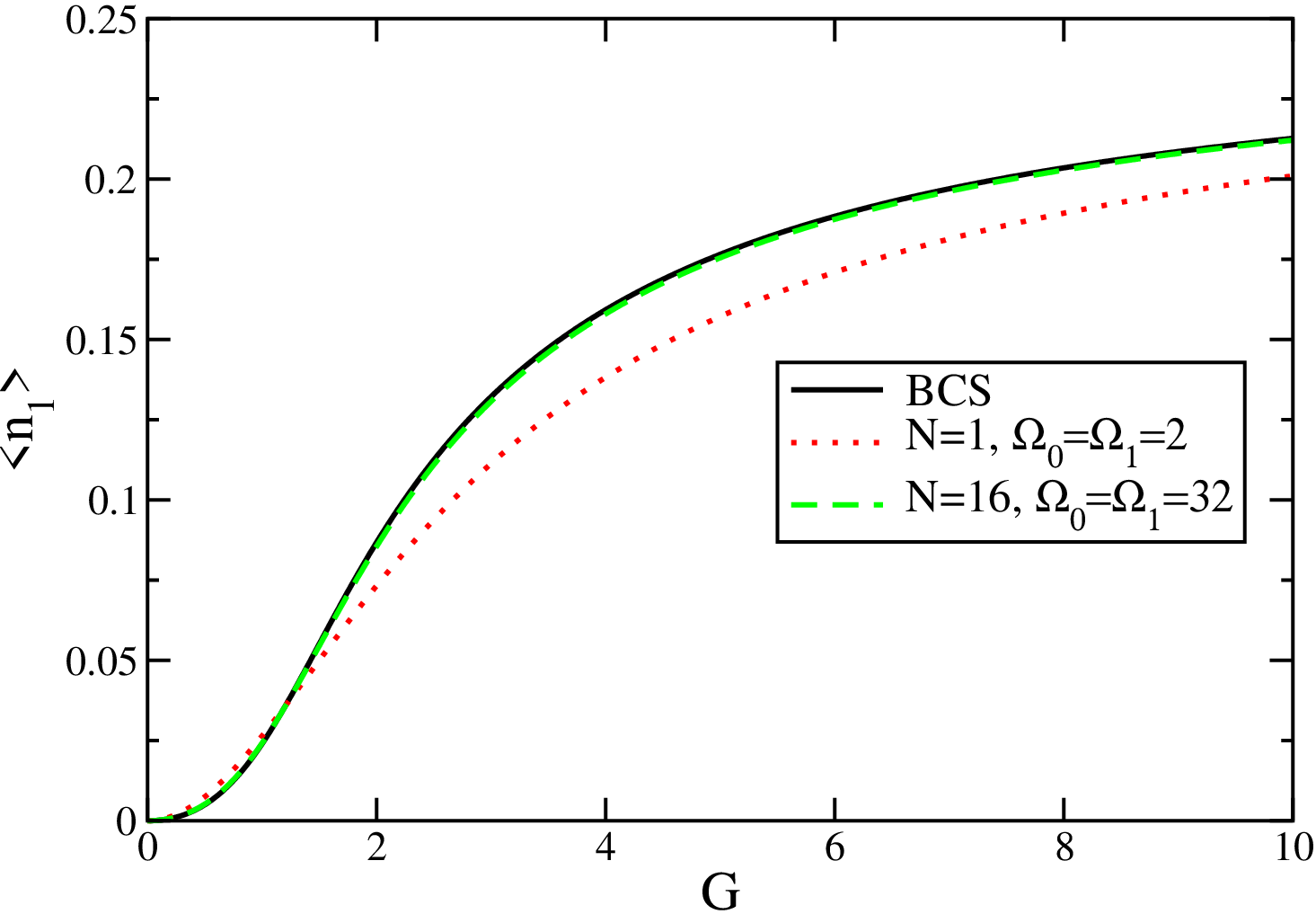}\\
  \caption{\label{fig:occup_one_quarter} (Color online) Average
    particle number of one of the upper levels as a function of $G$
    for various system sizes in the quarter-filled case $\delta=1/4$.
    The BCS Eqs.~(\ref{self_cons_eqns}) have real solutions for $G>0$
    in this case. The average particle number of one of the lower
    levels follows from
    $\langle\hat{n}_0\rangle=0.5-\langle\hat{n}_1\rangle$.}
\end{figure}

It is also interesting to study the system away from half filling. As
an example we now look at the case of quarter filling.  Again, we
assume that $\Omega_0=\Omega_1$ and to have a direct comparison to the
model at half filling, we chose the same system sizes as in the last
example. Because of particle-hole symmetry we have the following
relations between filling factors $\delta=N/(\Omega_0+\Omega_1)$ and
$1-\delta$.  For the average particle numbers
\begin{eqnarray}
  \langle\hat{n}_0\rangle_\delta&=&
1-\langle\hat{n}_1\rangle_{1-\delta}\;,\\
  \langle\hat{n}_1\rangle_\delta&=&
1-\langle\hat{n}_0\rangle_{1-\delta}\;,
\end{eqnarray}
and for the correlator
\begin{eqnarray}
  g(0,0^{\prime})_\delta&=&g(1,1^{\prime})_{1-\delta}\;,\\
  g(1,1^{\prime})_\delta&=&g(0,0^{\prime})_{1-\delta}\;,\\
  g(0,1)_\delta&=&g(0,1)_{1-\delta}\;.
\end{eqnarray}
In the following, we will consider the case of $\delta=1/4$ (which is
therefore equivalent to $\delta=3/4$). The average occupation of one
of the upper levels for this case is shown in
Fig.~\ref{fig:occup_one_quarter}. The occupation of one of the lower
levels is not shown, since it follows from $\langle \hat
n_0\rangle=2\delta-\langle \hat n_1\rangle$. We see that the
saturation at large interaction constant happens at larger $G$ than in
the half-filled case (\textit{c.~f.}~Fig.~\ref{fig:occup}). Note that
here the BCS solution always exists and is indistinguishable from the
exact solution already for 16 bosons.


\begin{figure}
  \includegraphics[width=\columnwidth,clip=true]{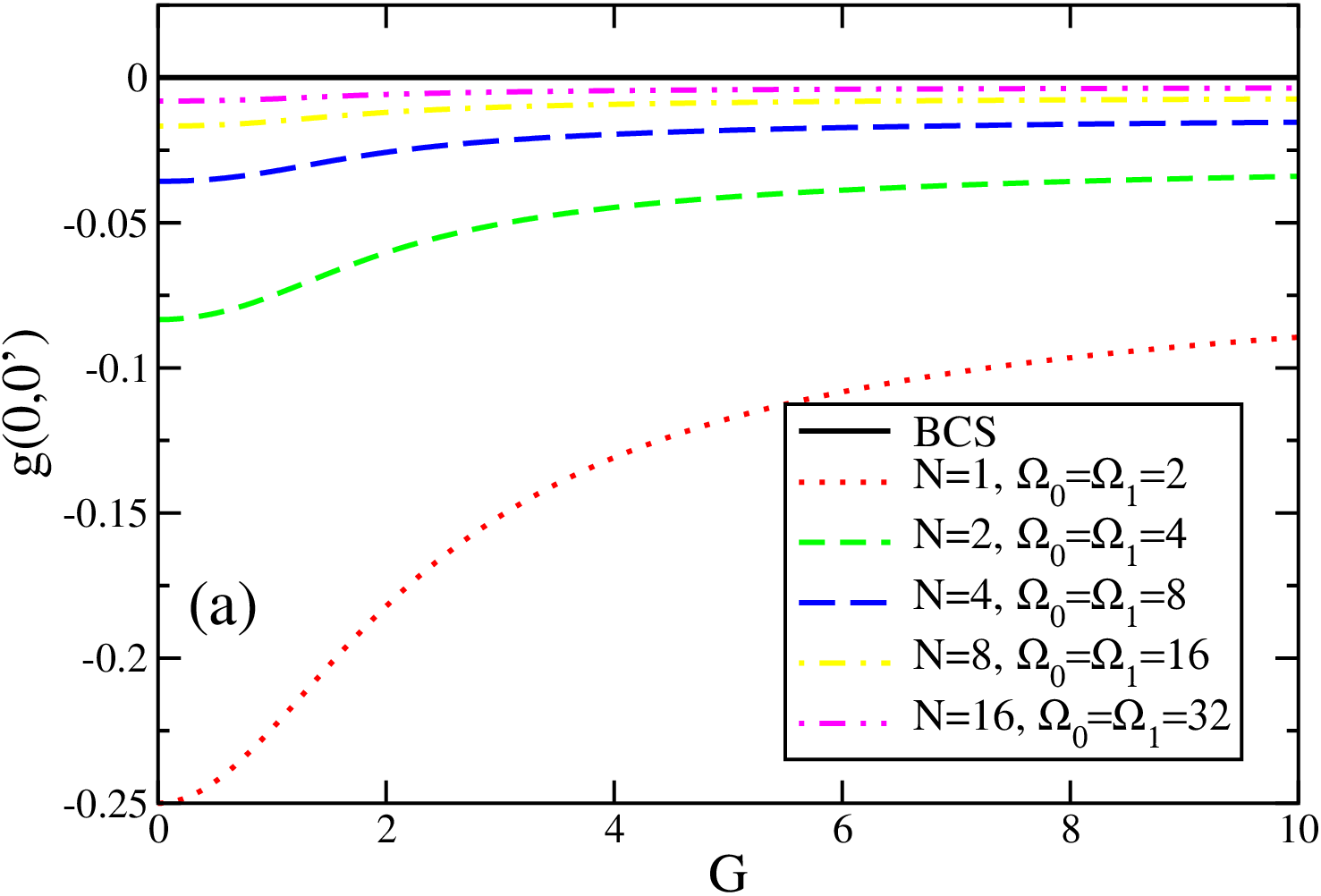}\\
  \includegraphics[width=\columnwidth,clip=true]{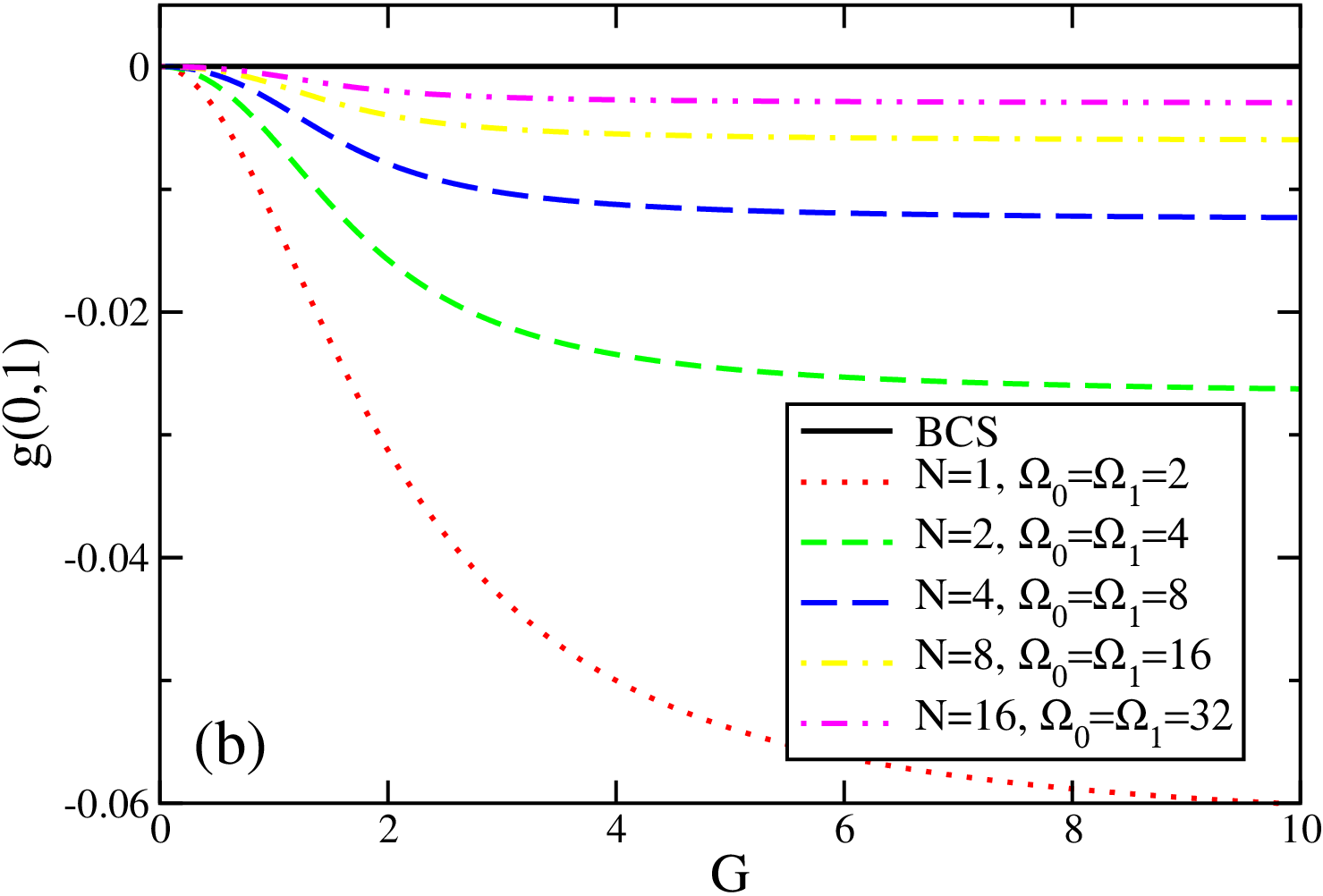}\\
  \includegraphics[width=\columnwidth,clip=true]{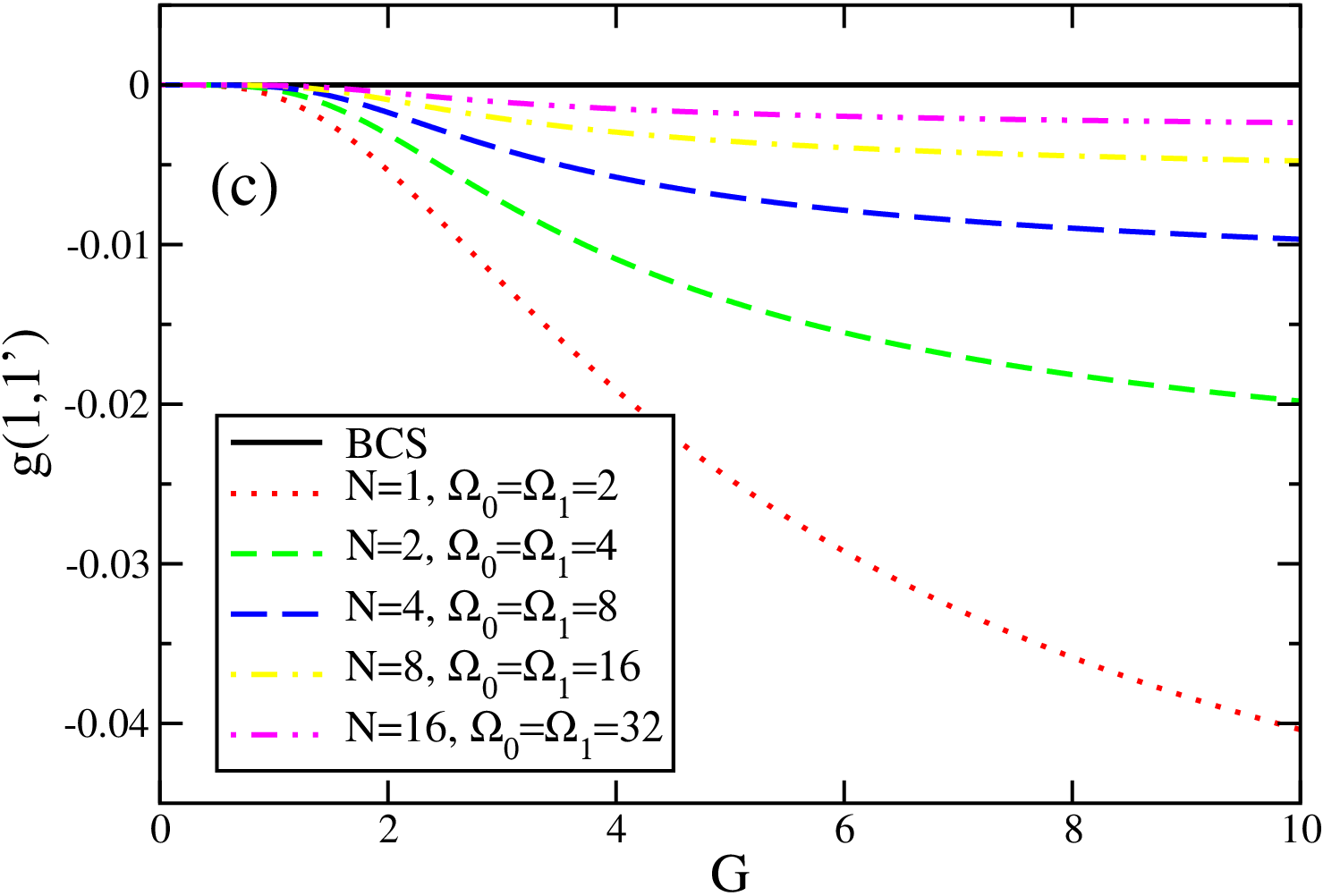}
  \caption{\label{fig:corr_one_quarter} (Color online)
  $g(0,0^{\prime})$, $g(0,1)$, and $g(1,1^{\prime})$ as a function of
  $G$ for various system sizes at quarter-filling $\delta=1/4$.}
\end{figure}

Figure~\ref{fig:corr_one_quarter} shows that all correlators are now
negative and different from each other and approach their limits in
the strongly interacting case more slowly than in the half-filled
case.  $g(0,0^{\prime})$ shows an interesting behavior for vanishing
interaction that is caused by the partially occupied lower energy band
allowing particles to change states among the lower levels. This leads
to a finite value also for $G\approx 0$ and a decay of the correlator
with increasing system size, see
Eq.~(\ref{eq:average_gll_asymptotic}). It is also remarkable that
$g(0,0^{\prime})$ is suppressed by increasing the interaction. Also,
in agreement with the perturbative results
Eq.~(\ref{eq:corr_auto_perturbative}) the effect of the interaction is
of second order in the interaction constant. The other correlators
show a similar behavior as in the half-filled case.

\section{Conclusion}

We have investigated exact particle-number correlations of ultracold
fermionic gases in a canonical Ansatz using the Richardson
solution. By means of a special configuration involving two degenerate
energy levels, correlation functions have been derived and evaluated
numerically for different mutual interactions between the atoms and
different system sizes. The particle numbers in different levels turn
out to be mostly anti-correlated revealing the fermionic origin of the
paired particles (the hard-core boson property). Approaching the
thermodynamic limit, those correlators decay to zero in the whole
interaction regime. This is in agreement with the predictions of BCS
theory. In the limit of strong interactions we were able to give
closed expressions for the correlations, which are also valid for the
general case of arbitrary level configurations. Due to the complex
algebraic structure of the Richardson solution, only a comparatively
special model could be investigated in this work. The discussion of
more general systems remains an open problem. Nevertheless, we believe
our predictions can be tested in tailored few-particle systems of
interacting fermions, e.~g. with atomic chips.

\begin{acknowledgments}
We would like to thank C. Schroll for useful discussions. This work
was financially supported by the Swiss National Science Foundation,
the NCCR Nanoscience, the European Science Foundation (QUDEDIS
network), and by the Deutsche Forschungsgemeinschaft within the SFB
513.
\end{acknowledgments}

\appendix*

\section{Perturbative calculation}

The Richardson solution and correlators can be found perturbatively in
the interaction constant. For $N\leq\Omega_0,~\Omega_1$ we find the
expression
\begin{widetext}
  \begin{eqnarray}
    g(0,0') & = & 
    -\frac{N(\Omega_0-N)}{\Omega_0^2(\Omega_0-1)} 
    \left[1-\left(\frac{G}{N}\right)^2\frac{\Omega_1(\Omega_0-N+1)}{2}\right]
    \nonumber\\ 
    & & - 
    \frac{N \Omega_1(\Omega_0-N+1)}{16\Omega_0^2(\Omega_0-1)}
    \left(\frac{G}{N}\right)^4\Big[2N(N-1)(5N-6)
    - (7N-2\Omega_0-7)(3N-\Omega_0-2)\Omega_0 
    \nonumber\\ 
    & & + 
    \big(13N(N-1)+7\Omega_0(2-3N+\Omega_0)\big)\Omega_1
    +2(N-\Omega_0)\Omega_1^2\Big]\,.
    \label{eq:corr_auto_perturbative}
\end{eqnarray}
For a half-filled band ($N=\Omega_0=\Omega_1$) the zeroth and the
second-order term vanish and the expansion to 6th order yields
\begin{eqnarray}
        g(0,0') \approx (\frac{1}{16N^2}-\frac{1}{8N^3})G^4 
        + (-\frac{3}{16N^2}+\frac{13}{32N^3}-\frac{3}{16N^4})G^6
        +\mathcal{O}(G^8)~.
\end{eqnarray}
The $G^4$ term is positive for $N>2$, but gets smaller for increasing $N$. For
increasing $G$ the 6th-order term takes over and leads to a negative
correlator in the end.
\end{widetext}

\end{document}